# Optimizing Service Orchestrations

Adam Barker
Department of Engineering Science, University of Oxford, UK.

*Abstract*—As the number of services and the size of data involved in workflows increases, centralised orchestration techniques are reaching the limits of scalability. In the classic orchestration model, all data passes through a centralised engine, which results in unnecessary data transfer, wasted bandwidth and the engine to become a bottleneck to the execution of a workflow.

This paper presents and evaluates the *Circulate* architecture which maintains the robustness and simplicity of centralised orchestration, but facilitates choreography by allowing services to exchange data directly with one another. *Circulate* could be realised within any existing workflow framework, in this paper, we focus on WS-*Circulate*, a Web services based implementation.

Taking inspiration from the Montage workflow, a number of common workflow patterns (sequence, fan-in and fan-out), input to output data size relationships and network configurations are identified and evaluated. The performance analysis concludes that a substantial reduction in communication overhead results in a 2–4 fold performance benefit across all patterns. An end-to-end pattern through the Montage workflow results in an 8 fold performance benefit and demonstrates how the advantage of using the *Circulate* architecture increases as the complexity of a workflow grows.

## I. INTRODUCTION

Efficiently executing large-scale, data-intensive workflows common to scientific applications must take into account the volume and pattern of communication. For example, in Montage [12] an all-sky mosaic computation can require at least 2–8 TB of data movement. Standard workflow tools based on a centralised enactment engine, such as Taverna [20] and OMII BPEL Designer [25] can easily become a performance bottleneck for such applications, extra copies of the data (intermediate data) are sent that consume network bandwidth and overwhelm the central engine. Instead, a solution is desired that permits data output from one stage to be forwarded directly to where it is needed at the next stage in the workflow. It is certainly possible to develop an optimised workflow system from scratch that implements this kind of optimisation. In contrast workflow systems based on concrete industrial standards offer a different set of benefits: they have a much larger and wider user base, which allows the leverage of a greater availability of supported tools and application components. This paper explores the extent to which the benefits of each approach can be realised. Can a standards-based workflow system achieve the performance optimisations of custom systems? And what are the trade-offs?

This paper explores these questions in the context of Web services, a widely-promoted standard for building distributed workflow applications based on a suite of simple standards (XML, WSDL, SOAP, etc.) designed to facilitate service interoperability. This paper does not address the performance limitations inherent in SOAP, an issue well addressed by other groups [9],[1],[6].

Workflow can be described from the view of a single participant using orchestration or from a global perspective using choreography. Web service orchestration enables Web services to be composed together in predefined patterns, described using an orchestration language and executed on an orchestration engine. Orchestrations can span multiple applications and/or organisations and result in long-lived, transactional processes. Services themselves have no knowledge of their involvement in a higher level application and therefore need no alteration before enactment. Importantly, Web service orchestrations are described from the view of a single participant (which can be another Web service) and therefore a central process always acts as a controller to the involved services. Orchestration languages explicitly describe the interactions between Web services by identifying messages, branching logic and invocation sequences. The Business Process Execution Language (BPEL) [22] is an executable business process modelling language and the current de-facto standard way of orchestrating Web services. BPEL has broad industrial support from companies such as IBM, Microsoft and Oracle, with concrete implementations.

Service choreography on the other hand is more collaborative in nature. A service choreography is a description of the externally observable peer-to-peer interactions that exist between services, therefore choreography does not rely on a central coordinator. A choreography model describes multi-party collaboration and focuses on message exchange; each Web service involved in a choreography knows exactly when to execute its operations and with whom to interact. A choreography definition can be used at design-time to ensure interoperability between a set of peer services from a global perspective, meaning that all participating services are treated equally, in a peer-to-peer fashion. The Web Services Choreography Description Language (WS-CDL) [14] is an XML-based language that can be used to describe the common and collaborative observable behaviour of multiple services that need to interact in order to achieve a shared goal. WS-CDL is a W3C Candidate Recommendation.

This paper presents the *Circulate* architecture, a hybrid solution that "eliminates the middle man" by adopting an orchestration model of central control, but a choreography model of optimised distributed data transport. Our architecture could be realised within any existing workflow framework, even custom systems. In this paper, we focus on a Web service based implementation for the evaluation. To explore the benefits of the hybrid approach for data-intensive applications, a set of workflow patterns and input-ouput relationships common to scientific applications (e.g. Montage) are used in isolation and combination. The performance analysis concludes that a substantial reduction in communication overhead results in a 2–4 fold performance benefit across all patterns. An end-to-end pattern through the Montage workflow demonstrates how the advantage of using the *Circulate* architecture increases when patterns are used in combination with another, resulting in a 8 fold performance benefit.

## II. SCIENTIFIC WORKFLOW PATTERNS

To identify data-centric scientific workflow patterns, the Montage application has been used. It is representative of a class of large-scale data-intensive scientific workflows. Montage constructs custom "science-grade" astronomical image mosaics from a set of input image samples [12]. The inputs to the workflow include the images in standard FITS format (a file format used throughout the astronomy community), and a "template header file" that specifies the mosaic to be constructed. The workflow can be thought of as having three parts, including re-projection of each input image to the coordinate space of the output mosaic, background rectification of the re-projected images, and co-addition to form the final output mosaic [8].

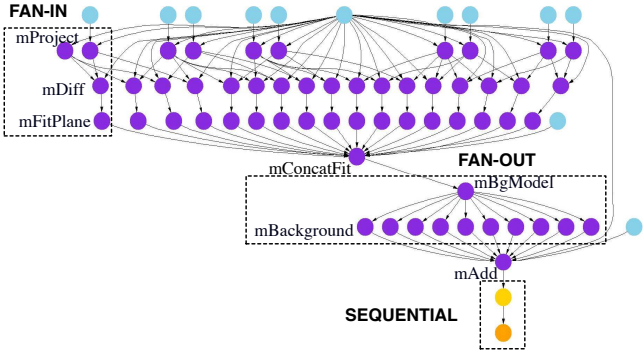

Fig. 1. Montage Use-case scenario.

A typical montage workflow is depicted in Figure 1. This workflow consists of the following six components (with input/output relationships listed):

1) mProject : reprojects a single image to the coordinate system defined in a header file (output = input)
2) mDiff/mFitPlane: finds the difference between two images and fits a plane to the difference image (output = 15–20 % of a typical image for each image triplet)
3) mConcatFit: a simple concatenation of the plane fit parameters from multiple mDiff/mFitPlane jobs into a single file (see 4)
4) mBgModel: models the sky background using the plane fit parameters from mDiff/mFitPlane and computes planar corrections for the input images that will rectify the background across the entire mosaic (output = a subset of inputs are output from mConcatFit and mBgModel)
5) mBackground: rectifies the background in a single image (output = input)
6) mAdd: co-adds a set of reprojected images to produce a mosaic as specified in a template header file (output = 70–90 % the size of inputs put together)

Montage illustrates several features of data-intensive scientific workflows. First, Montage can result in huge data flow requirements. For example, a small input file is 1.5 MB and a small Montage application can consist of hundreds of input files, a larger problem, 10K–100K image files, all input in the mProject phase. The intermediate data can be 3 times the size of the input data. And a big problem, e.g. an all-sky mosaic can result in 2-8 TB of data. Such a problem might be run daily. Second, Montage contains workflow patterns common to many scientific applications:

1) Sequence: This pattern involves the chaining of services together, where the output of one service invocation is used directly as input to another, i.e. serially (1:1 relationship). The data flows as a pipeline with no data transformations, e.g. mConcat → mBgModel.
2) Fan-in: Involves mapping multiple sources to a single sink (N:1 relationship), e.g. mDiff/mFitPlane → mConcatFit.
3) Fan-out: The reverse pattern of fan-in, data from a single source is sent to multiple sinks (1:N relationship), e.g. mBgModel → Background.

Large-scale scientific workflows such as Montage may also have significant computational requirements that must be considered in deployment. In this paper, we consider optimisation of workflow patterns as representative of a class of large-scale data-intensive scientific workflows. We focus only on the orchestrations and techniques required to reduce the cost of communication, assuming the computational resources for executing the workflow have been identified.

## III. HYBRID WORKFLOW ARCHITECTURE

The majority of workflow research has focused on service orchestration, where both control and data flow pass through a centralised server. There are a plethora of orchestration frameworks which will automate these tasks, examples of which can be found in the Business Process Modelling community through BPEL, in the Life Sciences through Taverna [20] and in the computational Grid community through Pegasus [8], Triana [21] and Kepler [18]. For a summary refer to [2]. Choreography, although an established concept is a less well researched and implemented architecture.

This paper proposes the *Circulate* architecture, based on centralised control flow, distributed data flow [16]. The *Circulate* architecture sits between a purely centralised solution (orchestration) and a purely decentralised solution (choreography). A centralised orchestration engine issues control flow messages to Web services taking part in the workflow, however enrolled Web services can pass data flow messages amongst themselves, like a peer-to-peer model. This model maintains the robustness and simplicity of centralised orchestration but facilities choreography by avoiding the need to pass large quantities of intermediate data through a centralised server.

*Circulate* is based on proxies, a lightweight, non-intrusive piece of middleware, which provides a gateway and standard API to Web service invocation. A proxy allows Web services to exchange data flow messages directly with one another thereby avoiding transferring them through a centralised server. Proxies are installed as "near" as possible to enrolled Web services; by near we mean preferably on the same Web server or network domain, so that communication between a proxy and a Web service takes place over a Local Area Network. Depending on the preference of an administrator, a proxy can be responsible for one Web service, 1:1 or many Web services, 1:N, illustrated by Figure 2.

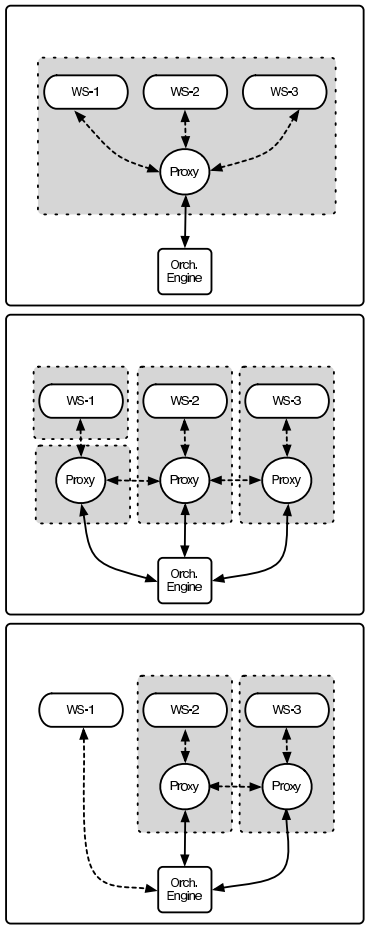

Fig. 2. 1:N (top), 1:1 (middle), mixed components (bottom).

Proxies themselves are exposed through a WSDL interface, allowing them to be built into workflows or higher level applications, such as any other Web service. As everything is exposed through a WSDL interface, this means that workflows can use a combination of proxies and vanilla Web services.

Proxies are controlled by a centralised orchestration engine which is executing an arbitrary workflow language, e.g. BPEL. However, only control flow messages are passed through the orchestration engine, larger data flow messages are exchanged between proxies in a peer-to-peer fashion, unless a proxy is explicitly told to do otherwise. Proxies exchange references to the data with the orchestration engine and pass the real data directly to where it is required for the next service invocation; this allows the orchestration engine to monitor the progress and make changes to the execution of a workflow. Unlike a pure choreography model, *Circulate* allows integration with centralised workflow systems making it easier to detect and handle failures.

Furthermore the architecture offers the following software engineering advantages:

- **Transition is non-disruptive:** The architecture can be deployed without disrupting current services and with minimal changes in the workflows that make use of them. This flexibility allows a gradual change of infrastructures, where one could concentrate first on improving data transfers between services that handle large amounts data.
- **Simplicity of deployment:** The proxy services can be installed without the need for writing any additional code. Configuration can be done remotely and dynamically. It simply requires the whereabouts of WSDL descriptions for any services that will be enabled through the proxy.
- **Non-intrusive deployment:** A proxy need not be installed on the same server as the Web service, and does not interfere with the current vanilla Web service as is the case with pure choreography models, e.g. WS-CDL. However, to gain more performance, the proxy should be as near as possible to the Web services it is enabling.

*A. Proxy Implementation and API*

The *Circulate* architecture is available as an open-source toolkit, WS-*Circulate* [3]; implemented using a combination of Java and the Apache Axis Web services toolkit [23]. Proxies are extremely simple to install and can be configured remotely, no specialised programming needs to take place in order to exploit the functionality. WS-*Circulate* is multi-threaded and allows several applications to invoke methods concurrently. A proxy has a thread pool and when that thread pool is full the request is placed on an input queue and dealt with in First In First Out (FIFO) order. Results from Web service invocations are stored at a proxy by tagging them with a UUID (Universally Unique Identifier) and writing them to disk. Proxies are made available through a standard WSDL interface, the Java representation of that interface is displayed in Figure 3. All methods are invoked by an orchestration engine except `stage`. A WS-*Circulate* proxy has the following methods:

- `invoke` is the primary proxy method and provides a gateway to Web service invocation. This method takes as input: details of the Web service to be invoked, including the location of a WSDL, portType and operation name, finally an array containing UUID references to data stored at the proxy; elements within the array must be in order as they would be used as input to the Web service. When this method is called the proxy retrieves the actual data the UUID references point to, using these data as input to the Web service invocation.

```java
public interface proxy {
   //Proxy CORE methods
   public String invoke(String wsdl, String port, String op_name, String[] params)
          throws InvocationParameterError, VariableNotFoundError, ServiceInvocationError;
   public String[] upload(Object[] params)
          throws InvocationParameterError;
   public boolean deliver(String proxy_wsdl, String[] dataToMove)
          throws VariableNotFoundError, ServiceInvocationError;
   public boolean stage(Hashtable dataToMove)
          throws ServiceInvocationError;
   public Object[] returnData(String[] dataToReturn)
          throws VariableNotFoundError;
   public boolean flushTempData(String [] dataToRemove)
          throws VariableNotFoundError;

   //Proxy ADMIN methods
   public void addService(String wsdl)
          throws ProxyAdminError;
   public void removeService(String wsdl)
          throws VariableNotFoundError;
   public String[] listOperations(String wsdl, String port)
          throws VariableNotFoundError;
   public String[] listOpParameters(String wsdl, String port, String op_name)
          throws VariableNotFoundError;
   public String[] listOpReturnType(String wsdl, String port, String op_name)
          throws VariableNotFoundError;
   public String[] listServices();
}
```

Fig. 3. WS-*Circulate* Proxy API

Any results are tagged with a UUID and written to disk at the proxy, this UUID is returned to the invoking application.

• **upload** provides functionality to upload data to a proxy which is required as input to a Web service invocation, i.e. if the service is the first within a workflow and is not reliant on data from services further up the chain. This method takes as input an `Object[]` that contains actual data to be uploaded. Elements within this array must conform to standard JAX-RPC supported types; the proxy will check this at runtime and exceptions will be thrown accordingly. Uploaded data are tagged with a UUID and written to disk, the corresponding UUIDs are returned to the invoking application.

• **deliver** sets up data movement between proxies, moving it closer to the source of a Web service invocation. The first input parameter is a `String` containing the location of the recipient proxy. Each element in the second input parameter, `String[]` represents one UUID reference to a blob of data stored at the proxy. Once invoked by an application the proxy will retrieve all data the UUID references point to and invoke the `stage` method on the recipient proxy. Currently data is moved using SOAP, however we are exploring the use of protocols such as GridFTP for large data transfer. An acknowledgement is returned represented as a `boolean`.

• **stage** is used to transfer a set of data from one proxy to another. This method is called from within the `deliver` method on the recipient proxy and moves the data to the recipient proxy. An acknowledgement is returned, represented as a `boolean`.

• **returnData** can be used to retrieve stored data from a proxy when it is needed on a user's desktop, e.g. to obtain the final results at the end of a workflow. Once invoked, the proxy iterates the input array (`String[]`), which contains UUID references to data, storing them in an `Object[]`. This array is then returned to the invoking application.

• **flushTempData** is a house keeping method and is called to remove data from a proxy which is no longer required for any workflow components. This method takes a list of UUID references to data, `String[]` and returns a `boolean`.

• **addService/removeService** is used to instruct a proxy to maintain a new Web service, adding the WSDL to its repository or remove it from a proxy's control. The input `String` represents the WSDL of a new service.

• **listOperations** given a WSDL and a port type this method returns a `String[]` where each element is the name of an operation.

• **listOpParameters** given a string containing a WSDL which the proxy maintains, the port type and the name of an operation this method returns a `String[]` containing the types expected as input to an operation.

• **listOpReturnType** returns the return type information (represented as a `String`) of an operation given a WSDL, port type and operation name.

• **`listServices`** is used to query a proxy about which Web services it is currently maintaining. This information is returned in a `String[]`; each element represents one WSDL.

Proxies throw the following exceptions:

• **`InvocationParameterError`** is thrown if the service details (used an input) are not maintained by a proxy or if the types and/or number of parameters used in an input array do not match the actual Web service interface that the proxy is to invoke.

• **`VariableNotFoundError`** is thrown if there are any references to a WSDL or data which cannot be found at a proxy.

• **`ServiceInvocationError`** will be thrown if there are any faults with the actual Web service invocation, e.g. network failure, time-out etc.

• **`ProxyAdminError`** is thrown if an application is trying to add a Web service which is already maintained by the proxy, or if the WSDL location is invalid.

### B. Example Application: Fan-in Pattern

Referring back to the Montage scenario, Figure 4 is a UML Sequence diagram illustrating how the fan-in pattern (discussed in Section II) is orchestrated using a standard centralised orchestration engine. In this pattern three `sources` are queried for data, these data are combined and used as input to a final `sink` service, which processes these data and returns a results set. Figure 5 illustrates the *Circulate* architecture applied to the same pattern. In our examples three source services are mapped to one sink, red arrowed lines added to each of the diagrams illustrate data movement, data sizes added to the each of the Figures are arbitrary and used for illustrative purposes only.

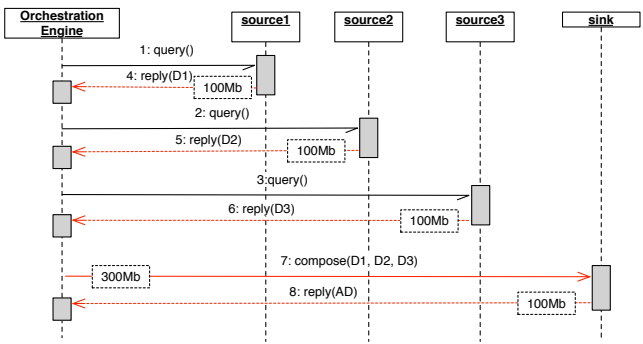

Fig. 4. UML Sequence diagram for orchestrated fan-in pattern

Using standard orchestration the query results (`D1`, `D2`, `D3`) from `source1-source3` pass through the centralised orchestration engine and are then used as input to the `sink` service, which analyses the data and returns the results (`AD`) back to the engine. Orchestration involves a total data flow of 700Mb.

With reference to Figure 5, in order to orchestrate the workflow using the *Circulate* architecture, the following process takes place. The first step in the workflow pattern involves making an invocation to the three source Web services

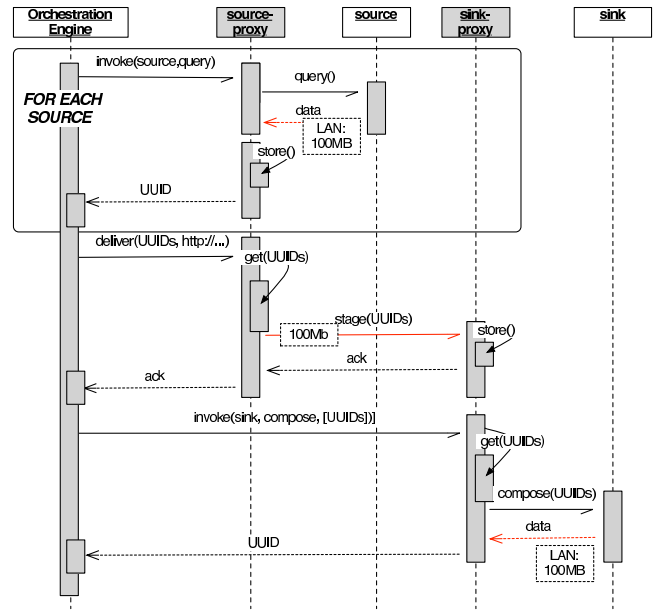

Fig. 5. UML Sequence diagram for fan-in pattern orchestrated using the *Circulate* architecture

`source1 - source3`. However instead of contacting the service directly, a call is made to a proxy (`source-proxy`) which has been installed on the same server as the Web service. This is achieved through the `invoke` operation passing the name of the Web service (`source`) and operation (`query`) to be invoked, along with any required input parameters. For readability portType details etc. have been omitted. The proxy spawns a new thread of control and invokes the `query` operation on the `source` service, passing in the necessary input parameters. The output from the service invocation, is passed back to the proxy, tagged with a UUID (for reference later, e.g. retrieval, deletion etc.) and stored; there is a requirement that the proxy has enough disk space to store the results. Instead of the proxy directly passing the data back to the orchestration engine, the UUID is returned. In a standard orchestration scenario the results of the Web service invocation would have first been moved to the orchestration engine and then moved to where they are needed at the `sink` Web service. However, as the proxy has been installed on the same server as the source data, it can be transferred locally between the proxy and the Web service and did not have to move over a Wide Area network, effectively saving a Wide Area hop. This process is repeated (either serially or in parallel) for `source2` and `source3` which could be served through the same proxy or an independent proxy.

The output from the Web service invocations are needed as input to the next service in the workflow, in this case the `sink` Web service. The orchestration engine invokes the `deliver` operation on the `source-proxy` passing in the three UUID references along with the WSDL address of the `sink-proxy`. Once the `source-proxy` receives the invocation it retrieves the stored data and transfers it across the network by invoking a `stage` operation on `sink-proxy`. The data is then stored at `sink-proxy` and if successful an acknowledgement message is sent back to `source-proxy` which is returned to the orchestration engine.

The final stage in the workflow pattern requires using the output from the first three services as input to the `sink` Web service. In order to achieve this the orchestration engine passes the name of the service (`sink`) and operation (`compose`) to invoke and the UUID references to the output data, which are required as input. The proxy then moves the data across the local network and invokes the `compose` operation on the `sink` service. The output is again stored locally on the proxy and a UUID reference generated and passed back to the orchestration engine. The orchestration engine can then retrieve the actual data from the proxy when necessary using the `returnData` operation.

Using the *Circulate* architecture the same quantity of data movement takes place, however only 300Mb of which is transfered through a Wide Area Network (i.e. proxy to proxy). The remaining 400Mb flows, ideally over a Local Area Network between the proxy and the subscribed Web service(s).

## IV. Performance Analysis

To verify our hypothesis we perform a set of performance analysis tests where the *Circulate* architecture is evaluated against a more traditional centralised control, centralised data flow orchestration engine.

### A. Experiment Description

Taking inspiration from the Montage workflow, we perform tests with the patterns common to many scientific applications (sequential, fan-in and fan-out) both in isolation and in a combination. Furthermore, we show best-case and worst-case performance of the *Circulate* architecture with respect to the location of the engine relative to the proxies. Throughout this paper we maintain the input to output data ratios discussed in Section II. With reference to Figure 6, the patterns have been configured as follows:

• **Pattern 1—Sequence:** The sequential pattern involves the chaining of services together, where the output of one service invocation is used directly as input to another. Once a service receives input data, its output is calculated by increasing the size of that input data by 20%, e.g. if the service receives 5Mb of data as input, 6Mb is returned as output. There is a snowball effect whereby the size of the data being transferred is increased after each service invocation. The configuration for this pattern on a fully centralised architecture is illustrated by phase 1 of Figure 6, and the configuration using the *Circulate* architecture is illustrated by phase 2 of Figure 6.

• **Pattern 2—Fan-in:** The fan-in pattern explores what happens when data is gathered from multiple distributed sources, concatenated and sent to a service acting as a sink. Multiple services are invoked with a control flow (no data is sent) message asynchronously, in parallel, a block of data is then returned as output. Once data has been received from all enrolled services it is concatenated and sent to the sink service as input, where 20% of that input is returned as output. The configuration for this pattern using a fully centralised architecture is illustrated by phase 3 of Figure 6 and the configuration using the *Circulate* architecture is illustrated by phase 4 of Figure 6.

• **Pattern 3—Fan-out:** This pattern is the reverse of the fan-in pattern, here the output from a single source is sent to multiple sinks. An initial service is invoked with a control flow message (again no actual data is sent), the service returns a block of data as output. These data are then sent, asynchronously in parallel to multiple services as input, each service returns as output the same size block of data it received as input. The configuration for this pattern using a fully centralised architecture is illustrated by phase 5 of Figure 6 and the configuration using the *Circulate* architecture is illustrated by phase 6 of Figure 6.

For each of the workflow patterns: sequence, fan-in and fan-out the time taken for the pattern to complete is recorded (in milliseconds) as the size of the input data (in Megabytes) is increased; for the sequential pattern this means the size of the file sent to the first service, for the remaining patterns this means the size of the input file returned by the first service. The number of services involved in each of the patterns range from 3 to 17, this takes into account the lower bound (mProject $\rightarrow$ mDiff) and upper bound (mFitPlane $\rightarrow$ mConcatFit) limits of the Montage workflow scenario discussed in Section II.

The configuration of our experiments mirror that of a typical workflow scenario, where collections of physically distributed services need to be composed into a higher level application. For each combination of input size, number of services and pattern type the experiment has been run independently 100 times over a cluster of distributed Linux machines. Wherever we report the time elapsed in milliseconds, 99% confidence intervals are included for each data point; some of these intervals are so small they are barely visible. Each line on the Figure 7, 8 and 9 displays the mean speedup ratio of each workflow pattern as the size of the input file increases. The mean speedup ratio is calculated by taking the average elapsed time (of 100 runs) for a vanilla (non-proxy, fully centralised) run of a workflow pattern and dividing it by the average elapsed time (of 100 runs) using the *Circulate* architecture. The number of services involved is independent of the ratio as we have taken the mean ratio for all combinations of services (i.e. running the experiment iteratively on 3 to 17 services) from our scaling experiments[1].

In order to explore locality, the placement of the orchestration engine is also taken into consideration, displayed on each graph are four sub-experiments, in descending order according to the graphs the following has been plotted:

• **Remote best-case:** The orchestration engine is entirely remote to the services/proxies it is invoking, by remote we mean that the orchestration engine has to connect over a Wide

---

[1]As an example, Appendix A displays the elapsed time of the sequence, fan-in and fan-out workflow patterns using 4 distributed services when the orchestration engine is remote.

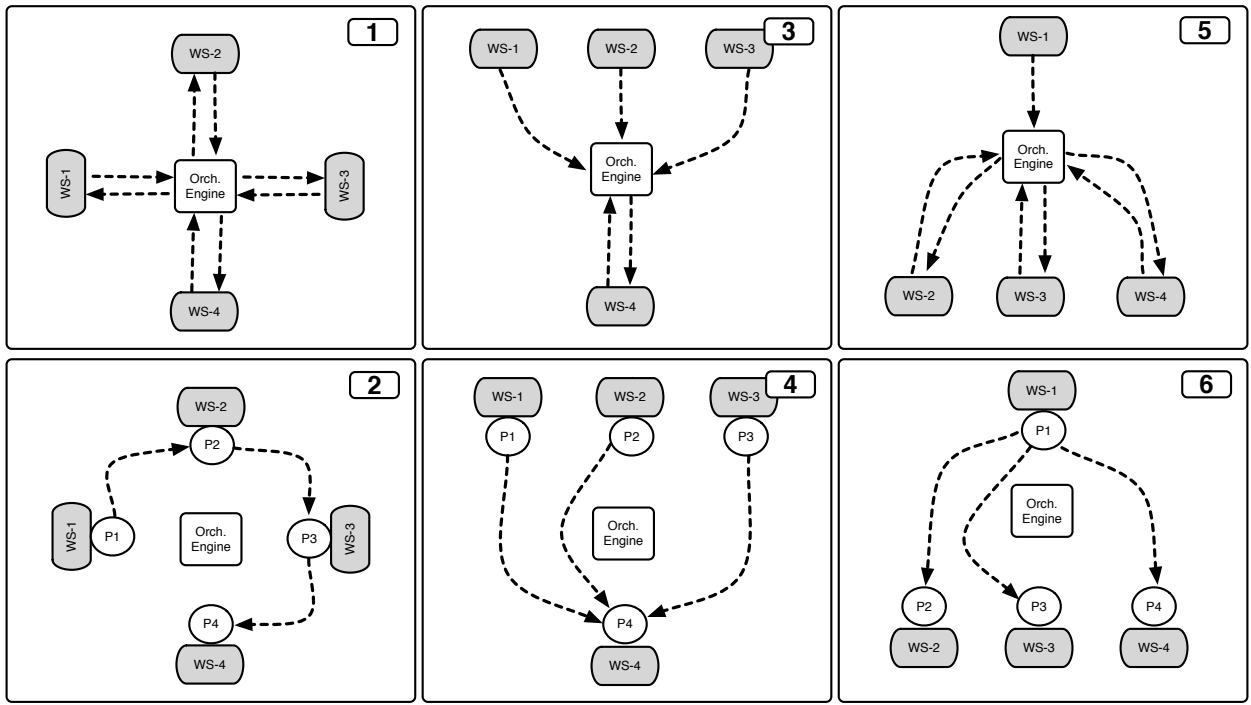

Fig. 6. Dataflow in the sequential (first column), fan-in (second column) and fan-out (third column) patterns for the centralised architecture using vanilla services (1,3,5) and the *Circulate* architecture (2,4,6). This example shows 4 services, all services are remote and all proxies are installed on the same server as the service they are invoking. Any control flow to orchestrate the services is omitted.

Area network. It is the best-case as the final results are stored on the proxy and not returned to the orchestration engine. The best-case scenario is realistic as often individual patterns form only a small piece of a larger workflow as highlighted by the Montage scenario.

• **Remote worst-case:** In this sub-experiment the orchestration engine is again remote but the final output data of the workflow pattern are not stored at the proxy but sent back to the orchestration engine.

• **Local best-case:** The orchestration engine is deployed locally (i.e. on the same network) as the services/proxies it is invoking. The best-case represents the scenario where the final output from the pattern execution is stored within a proxy.

• **Local worst-case:** The final sub-experiment represents the case where the orchestration engine is again local but the final output of the pattern is sent back to the orchestration engine.

The input and output data in all the experiments are Java byte arrays passed around using SOAP. To prevent the data processing from influencing our evaluation, it has not been accounted for in the performance analysis tests.

### B. Analysis of the Results

A collective summary of the performance analysis experiments is presented in Table I. Displayed on each row is the pattern type, the corresponding experiment configuration, i.e. where the orchestration is and how the proxy behaves (best/worst-case), along with the mean speedup ratio, standard deviation, minimum and maximum speedup ratios. The end-to-end pattern is discussed in Section IV-C

The performance analysis tests verify our hypothesis that when services are subscribed to the *Circulate* architecture the execution time of common, isolated workflow patterns significantly decreases.

The locality experiments confirm that the most dramatic benefit occurs when the orchestration engine is connected to the services/proxies through a Wide Area network. To quantify, the worst-case remote configuration, patterns saw an average performance benefit of between 2.03 and 2.83 times and in the best-case remote configuration patterns an average performance benefit of between 3.47 and 3.88 times, with the fan-in pattern showing the largest speedup.

A surprising result of our experimentation is that even when the orchestration engine is deployed on the same network as the services/proxies it is invoking (i.e. all communication is local) there is a benefit to using the *Circulate* architecture. In the worst-case local configuration patterns saw an average performance benefit of between 1.26 and 1.52 times and in the best-case local configuration patterns an average performance benefit of between 2.18 and 2.25 times.

To explain the results in relation to the *Circulate* architecture, when using a fully centralised approach the intermediate data have to make a costly hop back to the orchestration engine before being again sent across the network to be used as input to the next service in the workflow. However, using the *Circulate* architecture, intermediate data are stored at the proxy and sent directly to the next proxy which requires them as input, therefore for each input:output chain, one hop is avoided. In effect, this reduces the amount of intermediate data by 50%. This is, of course assuming that the proxy is installed as near as possible (i.e. on the same server or network) as the service it is invoking. This benefit is valid no matter where the orchestration is engine is placed, locally or remotely. Our locality experiments verify that even if workflows are orchestrated with locally deployed services the *Circulate* architecture speeds up the overall execution time of a workflow pattern. However, as the orchestration engines moves further away, the hop any intermediate data has to make increases in cost and the benefit of using the *Circulate* architecture increases accordingly. This explains why there is an increased benefit in the remotely deployed orchestration engine in relation to a locally deployed one. The difference in benefit is between 1.26 times and 1.70 times (mean remote-best − mean local-best across all patterns) in the best-case and between 0.74 times and 1.31 times (mean remote-worst − mean local-worst across all patterns) in the worst-case.

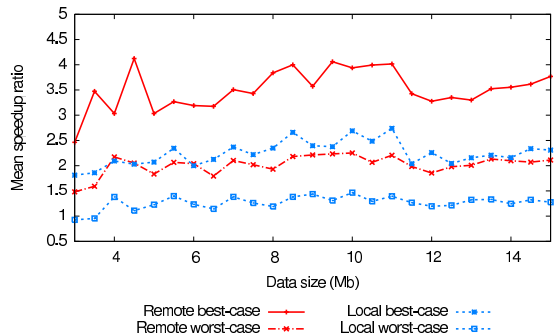

Fig. 7.  Sequential pattern mean speedup ratio, local vs. remote.

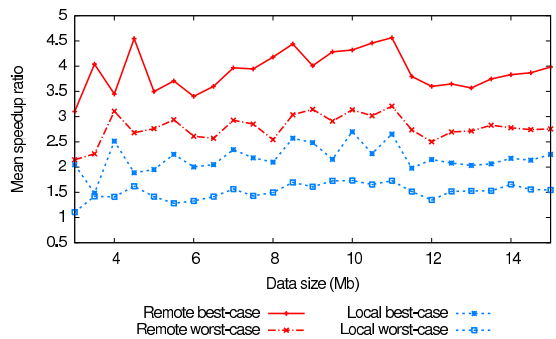

Fig. 8.  Fan-in pattern mean speedup ratio, local vs. remote.

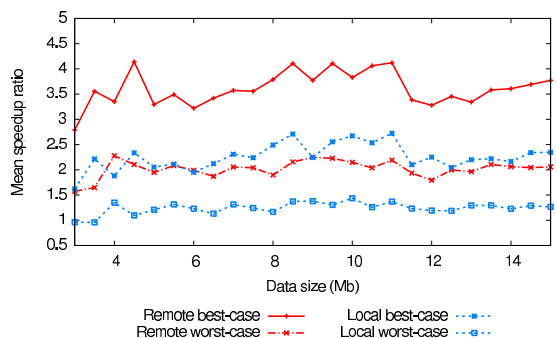

Fig. 9.  Fan-out pattern mean speedup ratio, local vs. remote.

The results (Figure 7, 8 and 9) confirm our intuition that the co-plots are bounded by remote best-case (best performance) and local worst-case (worst performance) for all patterns. The other cases lie in-between and their relative position depends on the specific pattern. The results also show that the relative speed up is mildly sensitive to data size. This can be explained as the speed up ratio depends on the relative amount of data sent and the relative network bandwidth for the local and non-local cases, both of which are approximately constant. The later may have some SOAP/HTTP/TCP dependencies which likely accounts for the small variation seen. However, the raw differential performance between the proxy and vanilla version does scale with data size (see Appendix A).

Although our experimentation is run at lower data sizes to Montage, patterns and input to output data relationships are maintained, this suggests that a similar performance benefit could be expected when scaling up the data injected into the workflow. Further experiments not discussed in this paper run over the PlanetLab [24] framework confirm that the ratios displayed Table I match those obtained from running the same experiments over an Internet scale network.

TABLE I
OVERVIEW OF THE PERFORMANCE ANALYSIS TESTS. BEST-CASE (BC) AND WORST-CASE (WC)

| Pattern | Config | Mean | Std Dev | Min | Max |
|---|---|---|---|---|---|
| Sequence | Local BC | **2.21** | 0.33 | 1.40 | 2.84 |
|  | Local WC | **1.29** | 0.14 | 0.93 | 1.51 |
|  | Remote BC | **3.47** | 0.54 | 1.83 | 4.41 |
|  | Remote WC | **2.03** | 0.18 | 1.48 | 2.28 |
| Fan-in | Local BC | **2.18** | 0.32 | 1.25 | 2.74 |
|  | Local WC | **1.52** | 0.19 | 0.97 | 1.81 |
|  | Remote BC | **3.88** | 0.53 | 2.23 | 4.97 |
|  | Remote WC | **2.83** | 0.27 | 2.14 | 3.41 |
| Fan-out | Local BC | **2.25** | 0.34 | 1.19 | 2.88 |
|  | Local WC | **1.26** | 0.13 | 0.96 | 1.49 |
|  | Remote BC | **3.61** | 0.51 | 2.13 | 4.94 |
|  | Remote WC | **2.07** | 0.21 | 1.57 | 2.63 |
| End-to-End | Remote WC | **8.18** | 0.94 | 5.58 | 9.86 |

## C. End-to-End Execution

Section IV-A and IV-B discussed workflow patterns in isolation, however the sequential nature of the Montage workflow suggests that the optimisations of different workflow patterns will have an end-to-end cumulative performance benefit, e.g. speeding up the time to perform mConcatFit will allow mBgModel to execute earlier, and so on. In order to verify this hypothesis a path through the Montage workflow was investigated, this end-to-end pattern is illustrated in Figure 10.

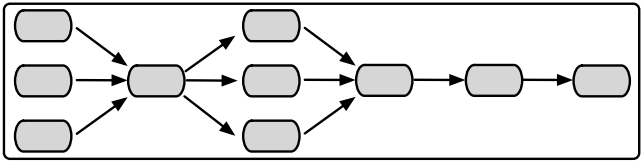

Fig. 10. An end-to-end workflow, with a fan-in, fan-out followed by a series of sequential operations.

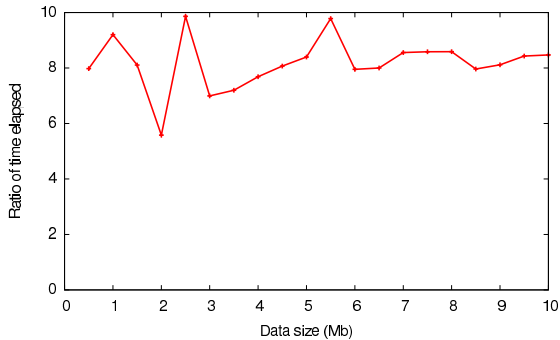

Fig. 11. End-to-end pattern.

This combination of patterns comprises of the following steps, firstly a fan-in pattern that asynchronously in parallel gathers data from 3 different services, the output of which is sent to a further service which returns 20% of the input data as output data. These data are then sent asynchronously in parallel to 3 services which each return the same volume of output data as they received as input. The output data is concatenated and sent through a further 2 services in sequence, each return 50% of the data they received as input. These input output data relationships mirror those found in the Montage scenario.

The end-to-end pattern displayed in Figure 10 is executed 100 times on on the *Circulate* architecture (using the worst-case, i.e. the final output data returns to the orchestration engine) and 100 times on a fully centralised orchestration engine with vanilla Web services, on both occasions the orchestration engine is remote. In Figure 11 the $x$-axis displays the input data size in Megabytes and the $y$-axis displays the ratio (vanilla elapsed time divided by proxy elapsed time) in milliseconds to complete the workflow. The end-to-end execution results in a mean speedup of 8.18 times using the proxy architecture, confirming our hypothesis that the performance benefit increases when isolated patterns are placed together to form a larger workflow. This sample end-to-end execution demonstrates the concept, however this combination pattern itself would only form a small part of larger scientific workflows, such as Montage.

## D. Break even Point

Invoking a proxy has an overhead in that a call is first made to a proxy, which invokes the service on the orchestration engines behalf, writes the result to disk and then returns a reference to that data. As the previous performance analysis tests demonstrate what occurs on relatively large data sizes, it is important to highlight what happens when dealing with Kilobytes instead of Megabytes of data in order to determine the break even point, i.e. when using a proxy is preferable over a vanilla service invocation.

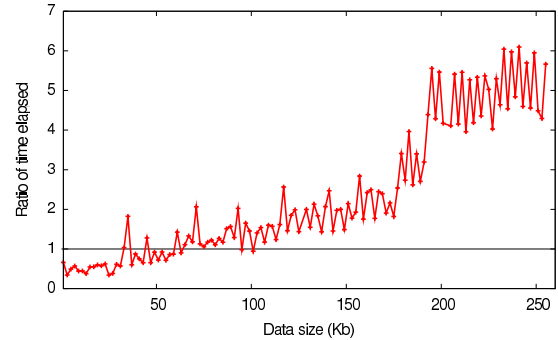

Fig. 12. The overhead of invoking a proxy.

Figure 12 displays the average time (as a ratio: vanilla elapsed time divided by proxy elapsed time) it takes to make a single invocation to a vanilla Web service and obtain the result vs. an invocation to a proxy that invokes the service on the orchestration engines behalf and returns a reference to its data. Results under the horizontal line indicate the vanilla approach is optimal, results over the line show a benefit of using the *Circulate* architecture. From the results we conclude that due to the overhead of the proxy, when dealing with input data sizes of less than ∼100K of data the *Circulate* architecture offers no performance benefit to vanilla Web services. Anything over ∼100K of data the proxy begins to speedup the execution time of the invocation. The *Circulate* architecture is suited to larger scale workflows (such as Montage) and not workflows where very small quantities of intermediate data are passed around between services, i.e. typical scenarios in business, such as transactions.

## V. RELATED WORK

This Section discusses all related work from the literature, spanning pure choreography languages, enhancements to widely used modelling techniques, i.e. BPMN, decentralised orchestration, data flow optimisation architectures and Grid toolkits.

### A. Choreography Languages

There are an overwhelming number of pure orchestration languages. However, relatively few targeted specifically at choreography.

The **Web Services Choreography Description Language (WS-CDL)** is the proposed standard for service choreography. However, WS-CDL has met criticism [4] through the Web services community. It is not within the scope of this paper to provide a detailed analysis of the constructs of WS-CDL, this research has already been presented [10]. However, it is useful to point out the key criticisms with the language: WS-CDL choreographies are tightly bound to specific WSDL interfaces, WS-CDL has no multi-party support, no formal foundation, no explicit graphical support and incomplete implementations.

**Let's Dance** [26] is a language that supports service interaction modelling both from a global and local viewpoint. In a global (or choreography) model, interactions are described from the viewpoint of an ideal observer who oversees all interactions between a set of services. Local models, on the other hand focus on the perspective of a particular service, capturing only those interactions that directly involve it.

**BPEL4Chor** [7] is a proposal for adding an additional layer to BPEL to shift its emphasis from an orchestration language to a complete choreography language. BPEL4Chor is a collection of three artifact types: participant behavior descriptions, participant topology and participant groundings.

### B. Techniques in Data Flow Optimisation

There are a limited number of research papers which have identified the problem of a centralised approach to service orchestration when dealing with data-centric workflows. For completeness, this Section presents an overview of a number of architectures.

The **Flow-based Infrastructure for Composing Autonomous Services** or FICAS [17] is a distributed data-flow architecture for composing software services into what the authors label mega-structures or workflow as it's more commonly known. Composition of the services in the FICAS architecture is specified using the Compositional Language for Autonomous Services (CLAS), which is essentially a sequential specification of the relationships among collaborating services. This CLAS program is then translated by the build-time environment into a a control sequence that can be executed by the FICAS runtime environment.

Although FICAS is an architecture for decentralised orchestration it does not deal directly with modern standards and is a prototype and proof of concept. The issue of Web services integration is not addressed, nor does it discuss how this architecture could be incorporated into an orchestration language such as the de-facto standard, BPEL. More importantly FICAS is intrusive to the application code as each application that is to be deployed needs to be wrapped with a FICAS interface. In contrast, our proxy approach is more flexible as the services themselves require no alteration and do not even need to know that they are interacting with a proxy. Furthermore our proxy approach introduces the concept of passing references to data around and deals directly with modern workflow standards.

**Service Invocation Triggers** [5], or simply triggers are also a response to the problem of centralised orchestration engines when dealing with large-scale data sets. Triggers collect the required input data before they invoke a service, forwarding the results directly to where the data is required. For this decentralised execution to take place, a workflow must be deconstructed into sequential fragments which contain neither loops nor conditionals and the data dependancies must be encoded within the triggers themselves.

The approach outlined by our paper and Service Invocation Triggers both rely on proxies to solve the problem of decentralised orchestration. While Triggers address the issue of decentralised control, to realise these benefits their architecture is based around a pure choreography model, which as discussed in this paper has many extra problems associated with it. Furthermore, before execution can begin the input workflow must be deconstructed into sequential fragments, these fragments cannot contain loops and must be installed at a trigger; this is a rigid and limiting solution and is a barrier to entry for the use of proxy technology. In contrast with our proxy approach, because data references are passed around, nothing in the workflow has to be deconstructed or altered, which means standard orchestration languages such as BPEL can be used to coordinate the proxies. Finally, Triggers does not deal with modern Web service standards.

In [19] the scalability argument made in this paper is also identified. The authors propose a methodology for transforming the orchestration logic in BPEL into a set of individual activities that coordinate themselves by passing tokens over shared, distributed tuple spaces. The model suitable for execution is called Executable Workow Networks (EWFN), a Petri nets dialect.

### C. Other Relevant Techniques

**Triana** [21] is an open-source problem solving environment. It is designed to define, process, analyse, manage, execute and monitor workflows. Triana can distribute sections of a workflow to remote machines through a connected peer-to-peer network.

**OGSA-DAI** [13] is a middleware product which supports the exposure of data resources on to Grids. This middleware facilitates data streaming between local OGSA-DAI instances.

**Grid Services Flow Language (GSFL)** [15] addresses some of the issues discussed in this paper in the context of Grid services, in particular services adopt a peer-to-peer data flow model. However, individual services have to be altered prior to enactment, which is an invasive and custom solution, something which is avoided in the *Circulate* architecture.

**Graph-forwarding** [11] is a technique applied to distributed Objects, allowing the results of an RPC to be forwarded to the next object to invoke instead of the invoking object. This technique is similar in nature but does not address the issues concerning service composition through workflow technology.

## VI. CONCLUSIONS

This paper presented the *Circulate* architecture for executing large-scale data-centric scientific workflows. Our architecture maintains the robustness and simplicity of centralised orchestration, but facilitates choreography by allowing services to exchange data directly with one another. Using Montage as a guide, a number of common workflow patterns and input-output relationships were evaluated in a Web services based framework. Although this paper discussed the *Circulate* architecture in a Web services context (WS-*Circulate*), it is a general architecture and can therefore be implemented using different technologies and integrated into existing systems. Furthermore the *Circulate* architecture is non-invasive to the Web services themselves.

Unlike the standard orchestration model, proxies can exchange data flow messages directly with one another avoiding the need to pass large quantities of intermediate data through a centralised server. The results indicate that substantial reduction in communication overhead results in a performance benefit of between 2.03 and 3.88 times. The advantage of using the *Circulate* architecture increases if isolated patterns are used in combination with another, the end-to-end pattern demonstrates an 8 fold performance benefit.

Future directions include evaluating the benefits of our approach within other workflow frameworks and in other network environments (e.g. wide-area, mobile) to assess the impact in different contexts. The analysis of additional applications to identify and evaluate other end-to-end workflow patterns is also planned. *Circulate* opens up a rich set of additional optimisations with respect to proxy deployment which will be evaluated in future work.

## REFERENCES

[1] Nayef Abu-Ghazaleh, Michael J. Lewis, and Madhusudhan Govindaraju. Differential Serialization for Optimized SOAP Performance. In *Proceedings of HPDC*, June 2004.

[2] Adam Barker and Jano van Hemert. Scientific Workflow: A Survey and Research Directions. In Roman Wyrzykowski and et al., editors, *Seventh International Conference on Parallel Processing and Applied Mathematics, Revised Selected Papers*, volume 4967 of *LNCS*, pages 746–753. Springer, 2008.

[3] Adam Barker, Jon B. Weissman, and Jano van Hemert. Orchestrating Data-Centric Workflows. In *The 8th IEEE International Symposium on Cluster Computing and the Grid (CCGrid)*, pages 210–217. IEEE Computer Society, May 2008.

[4] A. Barros, M. Dumas, and P. Oaks. A Critical Overview of the Web Services Choreography Description Language (WS-CDL). BPTrends Newsletter 3, 2005.

[5] Walter Binder, Ion Constantinescu, and Boi Faltings. Decentralized Ochestration of Composite Web Services. In *Proccedings of the International Conference on Web Services, ICWS'06*, pages 869–876. IEEE Computer Society, 2006.

[6] K. Chiu, M. Govindaraju, and R. Bramley. Investigating the Limits of SOAP Performance for Scientic Computing. In *Proccesings of the 11th International Symposium on High Performance Distributing Computing (HPDC)*, July 2002.

[7] G. Decker, O. Kopp, F. Leymann, and M Weske. BPEL4Chor: Extending BPEL for Modeling Choreographies. In *Proceedings of the IEEE 2007 International Conference on Web Services (ICWS 2007)*, pages 296–303, 2007.

[8] Ewa Deelman, Gurmeet Singh, Mei hui Su, James Blythe, A Gil, Carl Kesselman, Gaurang Mehta, Karan Vahi, G. Bruce Berriman, John Good, Anastasia Laity, Joseph C. Jacob, and Daniel S. Katz. Pegasus: A Framework for Mapping Complex Scientific Workflows onto Distributed Systems. *Scientific Programming Journal*, 13(3):219–237, 2005.

[9] K. Devaram and D. Andresen. Differential Serialization for Optimized SOAP Performance. In *Proceedings of PDCS*, November 2003.

[10] L. Fredlund. Implementing WS-CDL. In *Proceedings of the second Spanish workshop on Web Technologies (JSWEB 2006)*, 2006.

[11] Andrew. S. Grimshaw, Jon. B. Weissman, and W.T. Strayer. *Portable Run-time Support for Dynamic Object-Oriented Parallel Processing*, volume 14(2) of *Transactions on Computer Systems*. ACM, May 1996.

[12] J. C. Jacob, D.S. Katz, and et. al. The Montage Architecture for Grid-Enabled Science Processing of Large, Distributed Datasets. In *Proceedings of the Earth Science Technology Conference*, June 2004.

[13] Konstantinos Karasavvas, Mario Antonioletti, Malcolm Atkinson, Neil C. Hong, Tom Sugden, Alastair Hume, Mike Jackson, Amrey Krause, and Charaka Palansuriya. Introduction to OGSA-DAI Services. In *LNCS*, volume 3458, pages 1–12, 2005.

[14] N. Kavantzas, D. Burdett, G. Ritzinger, and Y. Lafon. Web Services Choreography Description Language (WS-CDL) Version 1.0. W3C Candidate Recommendation, 2005.

[15] S. Krishnan, P. Wagstrom, and G. von Laszewski. GSFL: A Workflow Framework for Grid Services. Technical report, Argonne National Argonne National Laboratory, 2002.

[16] David Liu, Kincho H. Law, and Gio Wiederhold. Analysis of Integration Models of Service Composition. In *Proceedings of Third International Workshop on Software and Performance*, pages 158–165. ACM Press, 2002.

[17] David Liu, Kincho H. Law, and Gio Wiederhold. Data-flow Distribution in FICAS Service Composition Infrastructure. In *Proceedings of the 15th International Conference on Parallel and Distributed Computing Systems*, 2002.

[18] B Ludascher and et al. Scientific Workflow Management and the Kepler System. *Concurrency and Computation: Practice and Experience*, 18(10):1039–1065, 2005.

[19] Daniel Martin, Daniel Wutke, and Frank Leymann. A Novel Approach to Decentralized Workflow Enactment. *EDOC '08. 12th International IEEE Conference on Enterprise Distributed Object Computing*, pages 127–136, 2008.

[20] T. Oinn, M. Addis, J. Ferris, D. Marvin, M. Senger, M. Greenwood, T. Carver, K. Glover, M. R. Pocock, A. Wipat, and P. Li. Taverna: a tool for the composition and enactment of bioinformatics workflows. *Bioinformatics*, 20:3045–3054, 2004.

[21] Ian Taylor, Matthew Shields, Ian Wang, and Roger Philp. Distributed P2P Computing within Triana: A Galaxy Visualization Test Case. In *17th International Parallel and Distributed Processing Symposium (IPDPS 2003)*, pages 16–27. IEEE Computer Society, 2003.

[22] The OASIS Committee. Web Services Business Process Execution Language (WS-BPEL) Version 2.0, 2007.

[23] Apache Axis: http://ws.apache.org/axis [16/12/2008].

[24] Planet Lab: http://www.planet-lab.org [16/12/2008].

[25] Bruno Wassermann and et al. Sedna: A BPEL-Based Environment for Visual Scientific Workflow Modelling. *Workflows for eScience - Scientific Workflows for Grids*, December 2006.

[26] Johannes Maria Zaha, Alistair Barros, Marlon Dumas, and Arthur ter Hofstede. Let's Dance: A Language for Service Behavior Modelling. In R Meersman and Tari Z, editors, *OTM Conferences (1)*, volume 4275 of *LNCS*, pages 145–162. Springer, 2006.

APPENDIX

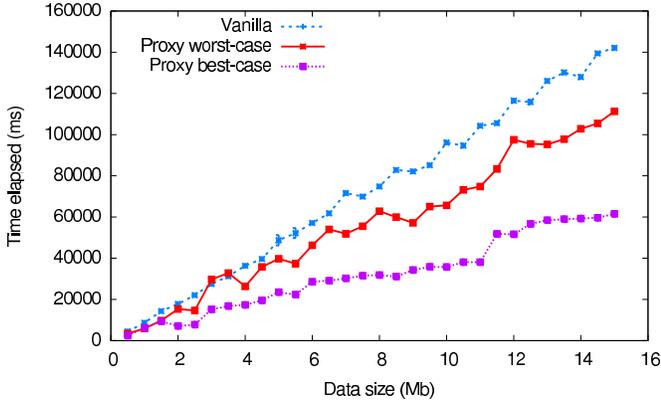

(a) Sequential pattern LOCAL orchestration engine

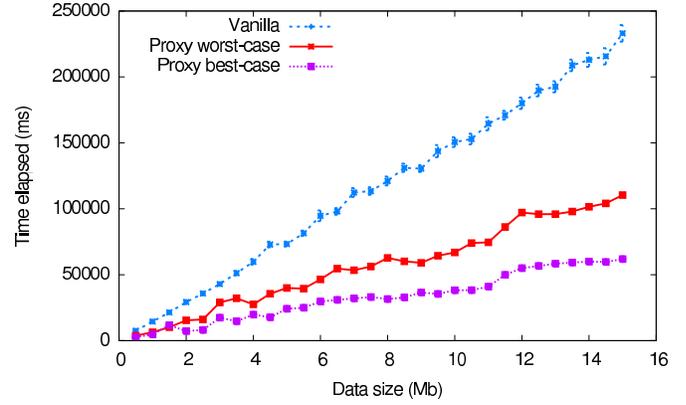

(b) Sequential pattern REMOTE orchestration engine

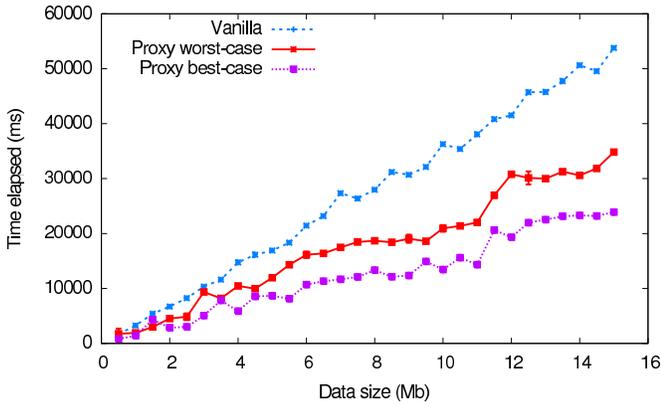

(c) Fan-in LOCAL orchestration engine

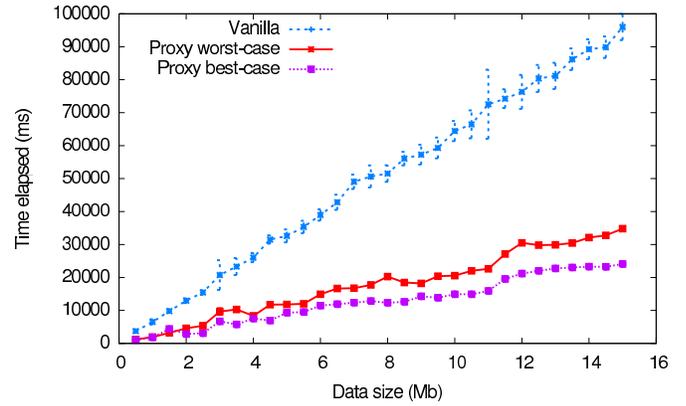

(d) Fan-in REMOTE orchestration engine

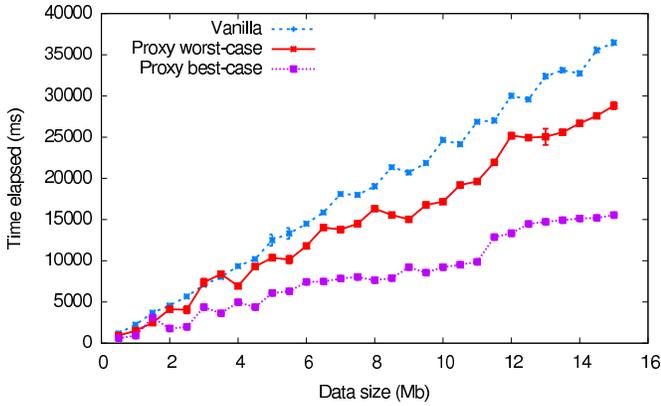

(e) Fan-out LOCAL orchestration engine

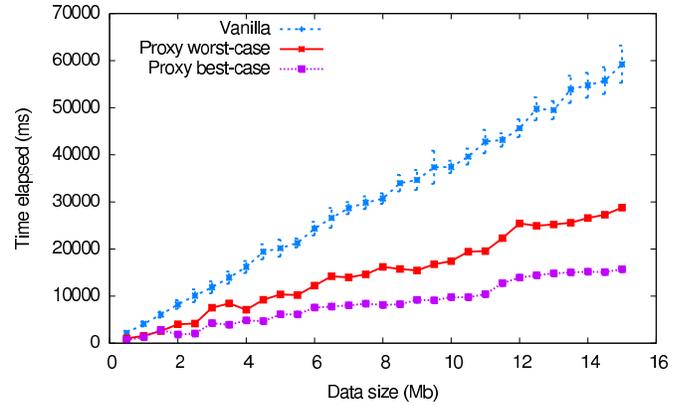

(f) Fan-out REMOTE orchestration engine

Fig. 13. An example experiment, using 4 services, recording the average time it takes for each pattern to complete as the size of the input data increases. The $x$-axis display the size of the initial input file in Megabytes (Mb) and the $y$-axis displays the elapsed time of the workflow pattern in milliseconds (ms). In 13(a), 13(c) and 13(e) the orchestration engine is locally deployed, in 13(b), 13(d) and 13(f) the orchestration is remotely deployed.